# Exploring the intrinsic energy resolution of liquid scintillator to approximately 1 MeV electrons

Y. Deng,[a] X. Sun,[b,c,1] B. Qi,[b,d] J. Li,[a] W. Yan,[b] L. Li,[b] H. Jiang,[f] C. Wang,[e] X. Cai,[b] T. Hu,[b] J. Fang,[b] X. Fan,[g] F. Gu,[a] J. Lv,[b] X. Ling,[a] G. Qu,[a] X. Qi,[a] L. Sun,[b] L. Zhou,[b] B. Yu,[b] Y. Xie,[b] J. Ye,[b,d] Z. Zhu,[b,d] Y. Zh,[a] G. Zuo[a,1]

[a] School of Nuclear Science and Technology, University of South China, Hengyang 421001, China
[b] Institute of High Energy Physics, Chinese Academy of Sciences, Beijing 100049, China
[c] State Key Laboratory of Particle Detection and Electronics, Institute of High Energy Physics, CAS, Beijing 100049, China
[d] School of Physical Sciences, University of Chinese Academy of Sciences, Beijing 100049, China
[e] National Engineering Research Center for Rare Earth Materials, General Research Institute for Nonferrous Metals, Beijing 100088, China
[f] School of Information Engineering, Jiangxi University of Science and Technology, Ganzhou 341000, China
[g] Beijing Key Laboratory of Passive Safety Technology for Nuclear Energy, School of Nuclear Science and Engineering, North China Electric Power University, Beijing 102206, China

E-mail: sunxl@ihep.ac.cn, guopzuo@163.com

**ABSTRACT:** We proposed a novel method for exploring the intrinsic energy resolution of a liquid scintillator (LAB + 2.5 g/L PPO + 3 mg/L bis-MSB) for approximately 1 MeV electrons. With the help of coincidence detection technology, single-energy electrons of $^{207}$Bi were effectively selected. With careful measurement and analysis of the energy resolution of a small liquid scintillator detector, the intrinsic energy resolution to 976 keV electrons was extracted to be 1.83% $\pm$ 0.06%. We used the wide-angle Compton coincidence (WACC) method to measure the luminescent nonlinearity of the liquid scintillator and found that it contributes only weakly to the intrinsic energy resolution of electrons. Such an unexpected large intrinsic energy resolution may come from fluctuations in energy transfer processes.

# 1 Introduction

Liquid scintillators are widely used radiation detection materials, especially for large-scale neutrino experiments [1-3]. The energy resolution of a scintillator detector is governed by several factors, including photon production fluctuation of the scintillator, photon collection of the detector, the Poisson component of photons to photoelectrons, electron multiplication of photomultiplier tubes (PMTs), and electronic noise. The photon production fluctuation of the scintillator is called the intrinsic energy resolution and is related to the incident particle types, energy, and scintillator materials. The intrinsic energy resolution and nonlinearity of inorganic scintillation detectors have been studied in detail by S.A Payne et al [4-7]. L. Swiderski et al used the wide-angle Compton coincidence (WACC) technique to measure the light yield nonlinearity and intrinsic energy resolution of some liquid and plastic scintillators [8]. A. Formozov et al determined that the intrinsic energy resolution of a liquid scintillator was 14.14% at 50 keV [9]. However, the intrinsic energy resolution of liquid scintillators at 1 MeV has not yet been investigated. In this work, we established a set of experimental and simulation methods for exploring the intrinsic energy resolution of liquid scintillators with approximately 1 MeV electrons.

For our experimental methods, we adopted state of the art coincidence detection techniques based on the internal conversion electron source $^{207}$Bi, in which single-energy electron events can be effectively selected through cascade events and energy spectrum analysis. A detailed analysis of the energy resolution of the detector is needed for extracting the intrinsic energy resolution [10]. As in Equation 1, the energy resolution of a specific liquid scintillator detector can be conveniently expressed as σ/E.

$$(\sigma/E)^2 = (\delta_{uniformity})^2 + (\delta_{Edep})^2 + (\delta_{int})^2 + (\delta_{st})^2 + (\delta_{system})^2 \quad (1)$$

where $\delta_{uniformity}$ is the non-uniformity of light collection caused by the difference in the light emitting position in the detector, including horizontal and vertical uniformity, $\delta_{Edep}$ is the fluctuation of deposited energy from particles emitted by the radioactive source, $\delta_{int}$ is the intrinsic energy resolution, $\delta_{st}$ is the statistical contribution of the Poisson component of photons to photoelectrons, $\delta_{system}$ is the contribution of the measurement system, including the gain of PMTs and electronic system. Therefore, the intrinsic energy resolution of the scintillator can be expressed as Equation 2.

$$\delta_{int} = \sqrt{(\sigma/E)^2 - (\delta_{st})^2 - (\delta_{system})^2 - (\delta_{Edep})^2 - (\delta_{uniformity})^2} \quad (2)$$

To determine the value of each term on the right side of the Equation 2, σ/E is the measured energy resolution of the detector, $\delta_{st}$ is determined by the number of photoelectrons and is expressed using Equation 3, $\delta_{system}$ is determined by using LaBr$_3$crystal calibration, single photoelectron calibration and long-term operation stability analysis, $\delta_{Edep}$ is given by Geant4 simulation, and $\delta_{uniformity}$, the horizontal uniformity measurement by moving the position of the radioactive source relative to the detector and the vertical uniformity is determined by Geant4 simulation.

$$\delta_{st} = \sqrt{1/N_{photoelectrons}} \quad (3)$$

We ultimately attempted to clarify the intrinsic energy resolution source by combining the luminescence nonlinearity with the energy deposition information obtained by simulation.

# 2 Experimental setups

## 2.1 Single-energy electron selection method and detectors setup

The internal conversion electron source $^{207}$Bi [11] was used in this experiment. There are three main gamma rays were emitted: $\gamma_1$(1770.2 keV), $\gamma_2$(1063.7 keV) and $\gamma_3$(569.7 keV). The two main single-energy internal conversion electrons near 1 MeV are e$_1$ (976 keV) and e$_2$ (1047 keV) and $\gamma_3$ radiation cascades with these conversion electrons. Therefore, a coincidence detector can tag this cascaded $\gamma_3$ to select the internal conversion electrons injected into the liquid scintillator detector. Another main function of this coincidence detector is to suppress Compton scattering events from $\gamma_2$ (1063.7 keV) rays in the liquid scintillator detector.

According to the Compton scattering Equation 4, when the scattering angle $\theta > 160°$ the

Compton electron energy is $E_e > 852$ keV, which can affect the 1 MeV internal conversion electron peaks. In principle, as long as the coincidence detector is sufficiently large, these large-angle Compton scattering events will be recorded. Furthermore, by analyzing the energy spectrum of the coincidence detector, a narrow energy cut of the $\gamma_3$ full energy peak can eliminate most of the large-angle scattering events.

$$E_e = \frac{(h\nu)^2(1-\cos\theta)}{m_e c^2 + h\nu(1-\cos\theta)} \quad (4)$$

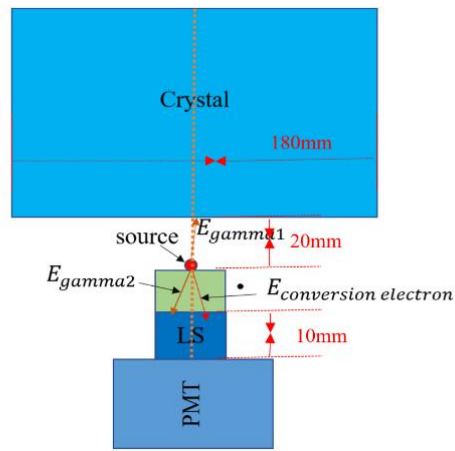

(a)

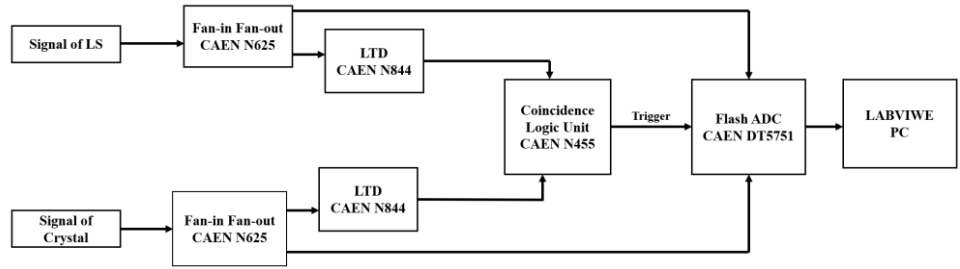

(b)

Figure 1. Experimental setup scheme for measuring the energy resolution of liquid scintillator (a) and experimental flowchart (b).

Based on this single-energy electron selection method, an experimental device was established, as shown in Figure 1. The liquid scintillator detector was a cylindrical quartz container with a diameter of 35 mm and height of 25 mm. The side of the container was wrapped with Teflon reflective film and the bottom was coupled to the PMT (CR160, Hamamatsu) using silicone grease. The container had a 10 mL liquid scintillator (LAB + 2.5 g/L PPO + 3 mg/L bis-MSB) and the liquid level was 10 mm high. The upper mouth of the container was covered with a flat Teflon reflective film, and there was a small hole with a diameter of 5 mm in the center of the reflective film to place the $^{207}$Bi electron source (3μ Ci).

The coincidence detector was a CsI(Na) crystal detector with crystal size 18 cm × 18 cm × 30 cm. The side of the crystal was wrapped with Teflon reflective film and encapsulated in an oxygen-free copper shell with quartz glass windows at both ends. Four PMTs (XP-5382) were coupled to one end of the crystal with silicone grease to read the photons signals, whereas the other end was covered with ESR reflective film. The 4-channel PMT signals used the fan-in-fan-out (N625) summation to form a one-channel detector signal. The coincidence detector was vertically arranged above the liquid scintillator detector, and its lower end was 20 mm from the upper end of the liquid scintillator detector. The radioactive source was placed between the two detectors at a distance of 15 mm from the liquid scintillator level and 20 mm from the crystal. The two detectors and the electron source were arranged coaxially. The coincidence signals of the liquid scintillator and the coincidence detector was used to trigger the data acquisition system (DT5751).

When the coincidence detector recorded a $\gamma_3$ (569.7 keV) event, two main cases were recorded by the liquid scintillator detector. 1) Small-angle Compton electrons produced by $\gamma_2$ rays; and 2) internal conversion electrons $e_1$ and $e_2$. The large-angle Compton scattering events of $\gamma_2$ in the liquid scintillator detector deposited energy in the coincidence detector, causing the total energy in the coincidence detector to deviate from $\gamma_3$ (569.7 keV). Therefore, matching the energy selection of 569.7 keV in the coincidence detector allowed for suppression of the large-angle Compton scattering event of $\gamma_2$ in the liquid scintillator detector. The intensity of branching ratio of $\gamma_1$(1770.2 keV) was relatively small, and the coincidence detector had the same suppression effect on the Compton scattering events of $\gamma_1$ in the liquid scintillator detector, therefore, its impact is negligible here.

## 2.2 Calibration method and setup for experimental system and horizontal uniformity contributions

The contribution of the test system was mainly introduced by the uniformity of the PMT photocathode, the fluctuation of the photoelectron multiplication and the readout electronics. It is difficult to directly measure the contributions of these items. Here, we adopted the method of replacing liquid scintillators with LaBr$_3$ crystals with better energy resolution to provide the upper limit of the test system contribution. The crystal size was 1 inch in diameter and 1 inch in height, which is similar to those of the liquid scintillator vessel. The crystal was placed in the same position of the PMT window, as shown in Figure 2a. The crystal is wrapped with a Teflon reflective film, except for the surface coupled with the PMT. The resulting photon collection effect was similar to that of the liquid scintillator. The electron source was shielded by the package shell of the crystal; we therefore chose 662 keV $^{137}$Cs as the excitation source. To simplify the discussion, all contributions except for the contribution of photoelectron statistics are attributed to the test system. Therefore, we can obtain the upper limit of the contribution of the measurement system $\delta_{system} \leq \sqrt{(\sigma/E)_{LaBr3}{}^2 - (\delta_{st_{LaBr3}})^2}$. The calibration of the single photoelectron was realized by using the LED light source driven by the pulse generator. The calibration of the long-term running stability of the system was obtained by analyzing the energy spectrum of the liquid scintillator detector in time segments. These two items are listed separately.

The contribution to the uniformity of the liquid scintillator detector was due to the different collection efficiencies of photons emitted at different incident positions within the liquid scintillator. For horizontal uniformity calibration, as shown in Figure 2b, we placed the $^{207}$Bi electron source with 5 mm diameter and 3 mm thickness lead collimation holes at different positions in the diameter direction of the liquid scintillator detector. Teflon reflective film was attached to the underside of the lead plate and covered on top of quartz container and there was a hole of the same size in the position of the lead collimation hole. Eight points were then measured at 5 mm steps. The contribution of horizontal uniformity can be thus expressed by the following Equation 5.

$$\delta_{uniformity} = \sigma/mean \qquad (5)$$

Here, the mean represents the average value of the ADC peak value of the eight test points, and $\sigma$ represents the standard deviation.

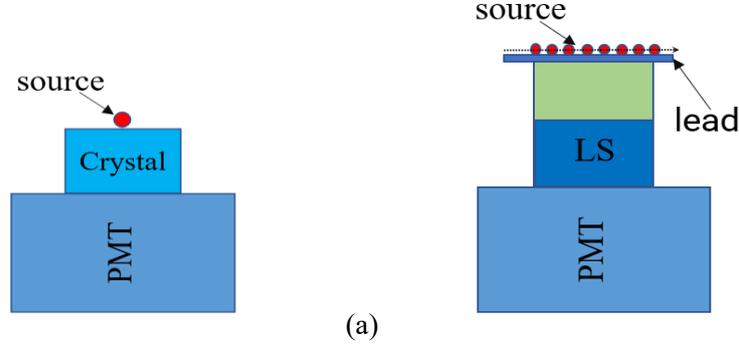

(a)                  (b)

Figure 2. (a) Experimental setup for measuring the system contribution. (b) Experimental setup for measuring the horizontal uniformity contribution.

## 2.3 Non-linearity of the light yield measurement setup

We used the WACC [12-14] method to measure the luminescence nonlinearity of a liquid scintillator. The advantage of the WACC technique is that it enables precise measurements using a small experimental setup with only one reference detector and weak radioactive γ-ray sources, without any collimators. The experimental setup is illustrated in Figure 3. It mainly includes a radioactive source ($^{137}Cs$), the same liquid scintillator detector, a high-purity germanium (HPGe) detector (CANBERRA's BE2020), and a shielded lead block. The investigated liquid scintillator detector was placed at a distance of 3 cm from the HPGe detector, and a 5 cm-thick lead brick was placed between the radioactive source and the HPGe detector. The distance and angle between the radioactive source and the liquid scintillator detector can be adjusted to obtain events with different scattering angles. The superb energy resolution of the HPGe detector provides a possibility to reconstruct the energy of Compton electrons in liquid scintillator $E_e$ directly from the difference between initial energy $E_\gamma$ (662 keV) and energy deposed in HPGe detector $E_\gamma^{'}$, that was $E_e = E_\gamma - E_\gamma^{'}$. The ratio of ADC output of liquid scintillator to deposited energy $ADC/E_e$ characterizes the nonlinearity of the light yield.

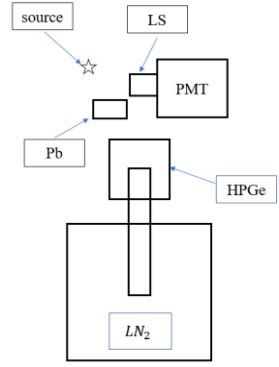

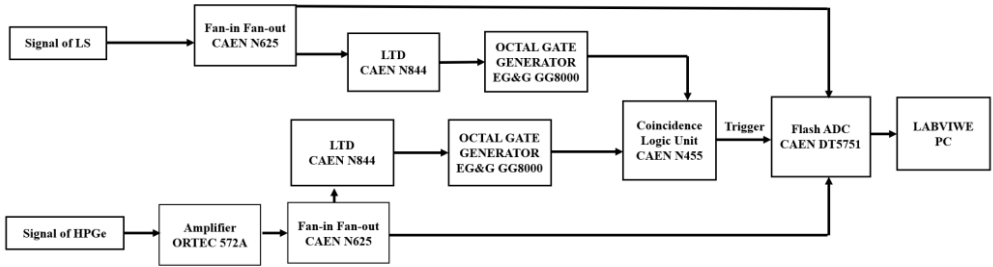

(b)

Figure 3. Experimental setup for measuring the non-linearity of light yield of liquid scintillator (a) and experimental flowchart (b).

# 3 Monte Carlo simulation

Geant4 is an object-oriented C++ program package developed by the European Organization for Nuclear Research (CERN), which is mainly used for Monte Carlo simulations of nuclear physics and high-energy physics experiments. Geant4 provides a complete range of cross-sectional data on the interaction between particles and matter, which can simulate the transport process of all known particles in materials. Here, we used Geant4 (version 10.2) to track the electrons, gamma rays of $^{207}$Bi and photons in the detectors. The simulation of electron and gamma deposition energy in liquid scintillator included physical processes of multiple scattering, ionization, bremsstrahlung and positron annihilation. Simulation of optical processes included absorption, scattering and reflection. The detector geometry was constructed based on the real experimental device. The main goals of the simulation were as follows:

(1) To understand the scattering process of gamma rays of $^{207}$Bi between the liquid scintillator and coincidence detectors and the degree of suppression of the large-angle Compton scattering events using coincidence measurement technology.

(2) To track the energy deposition process of electrons and gamma rays in the liquid scintillator detector to obtain the energy deposition spectrum as the input spectrum for fitting the experimental data.

(3) To track the energy deposition process of electrons in the liquid scintillator and obtain the initial electron energy and deposition energy of each step. Further, to convolve the deposited energy of each step to the light yield and sum to obtain the total light yield of an event. Finally, to study the contribution of the step distribution and the nonlinearity of the liquid scintillator to the intrinsic energy resolution.

(4) To track the propagation of photons within the detector and get uniformity of photon collection. A point light source and a simplified detector geometry that does not contain a radioactive source were used, each event generated 10,000 photons at a 4π solid angle.

Table 1. Optical simulation parameters.

|  | Liquid scintillator | Teflon film | Quartz glass | PMT glass |
|---|---|---|---|---|
| Refractive index | 1.498 | / | 1.47 | 1.47 |
| Absorption length (m) | 77 | / | 50 | / |
| Scattering length (m) | 27 | / | / | / |
| Reflectivity (%) | / | 95 | / | / |

# 4 Results

## 4.1 Energy spectrum of $^{207}$Bi

The two-dimensional scatter plots of the liquid scintillator detector and coincidence detector for the coincidence measurement are shown in Figure 4. Here, the X-axis represents the energy deposition in the liquid scintillator detector and the Y-axis represents the energy deposition in the CsI(Na) crystal detector. By selecting the $\gamma_3$ (mean($\gamma_3$) $\pm 0.5\sigma$) events in the coincidence detector, the corresponding energy spectrum of the liquid scintillator detector is shown in Figure 5a. Conversely, Figure 5b shows the energy spectrum of the liquid scintillator detector without event selection in the coincidence detector. The event selection strategy of the coincidence detector was successful and the Compton scattering events of gamma rays were effectively suppressed. The Geant4 simulation provided more obvious results. As shown in Figure 6, the event selection obviously suppressed the scattering events, causing the Compton edge to shift to the left.

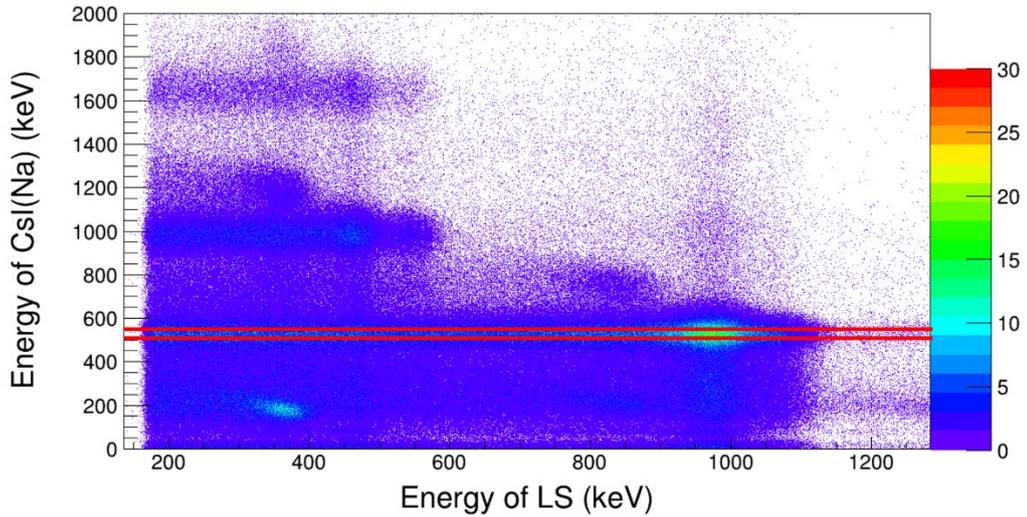

Figure 4. Two-dimensional scatter plots of the liquid scintillator detector and the coincidence detector for the coincidence measurement.

Figure 5a shows the double Gaussian fitting that was performed on the electron peak according to the energy and branch ratio of $e_1$ (976 keV) and $e_2$ (1047 keV), and the energy resolution of the electron with an energy of 976 keV is obtained as $\eta(\sigma/E) = 3.35\% \pm 0.10\%$. In contrast, Figure 5b is the energy spectrum without events selection, and the fitting resolution of $e_1$ (976 keV) is 3.57% $\pm$ 0.05%. In fact, the energy spectrum of $e_1$ can be extracted and fitted separately through simulation, which can eliminate the other remaining interference like associated X-rays and Auger electrons, see Section 4.2.

With the calibration of the PMT single photoelectron [15], it can be known that the number of photoelectrons obtained by the PMT is $N_{photoelectron}$ = 3266 ± 11. According to Equation 3, the contribution of the statistical fluctuation of photoelectrons to the energy resolution is calculated as $\delta_{st}$ = 1.75% ± 0.00%.

The influence of the cut width of $\gamma_3$ (569.7 keV) on the energy resolution of the liquid scintillator detector is shown in Table 2. The stricter the energy selection of 569.7 keV, the better the resolution of the corresponding liquid scintillator detector, but statistical reduction will eventually lead to a poor fit, worsening the energy resolution. In fact, the energy resolution around 0.5σ had stabilized.

Table 2. The influence of energy cutting range of $\gamma_3$ on energy resolution.

| Energy cutting range of $\gamma_3$ | Fitted interval (keV) | Results |
| --- | --- | --- |
| mean($\gamma_3$) ± 0.1σ | 940-1110 | 4.00% ± 0.39% |
| mean($\gamma_3$) ± 0.5σ | 940-1110 | 3.35% ± 0.10% |
| mean($\gamma_3$) ± 1.0σ | 940-1110 | 3.36% ± 0.08% |
| mean($\gamma_3$) ± 2.3σ | 940-1110 | 3.57% ± 0.05% |

Figure 5a shows that a platform structure was still present on the left side of the electron peak. It is known from simulation that this platform was mainly caused by the scattering events of electrons on the platinum metal substrate of the radiation source and the small-angle Compton scattering events.

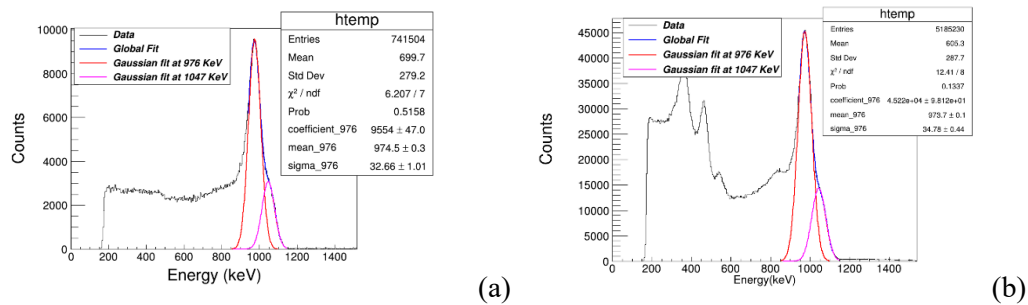

Figure 5. Energy spectrum of liquid scintillator with events selection (a) and without events selection (b).

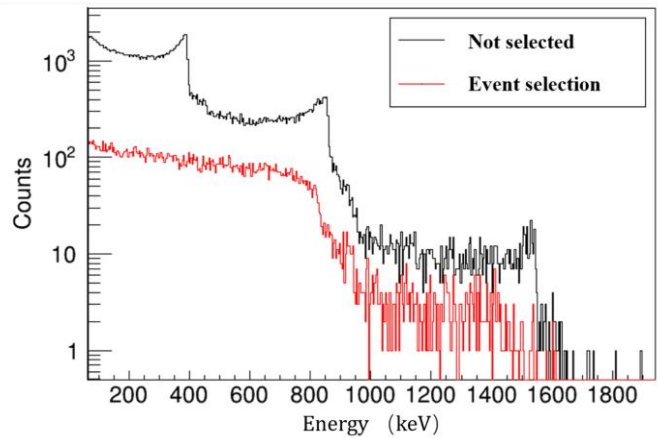

Figure 6. Simulation of the coincidence measurement of γ-ray energy deposition in the liquid scintillator; the black curve represents events without energy cut in the coincidence detector,

whereas the red curve represents events with energy cut in the coincidence detector.

By analyzing the total data in segments, the variation of the peak position of $e_1$ with time is shown in Figure 7, the data acquisition time of each segment was about 1 hour. Variation of system stability was less than 0.6%, the resulting system long-term operating stability $\delta_{stability}$ contributed 0.16% $\pm$ 0.03% to the energy resolution.

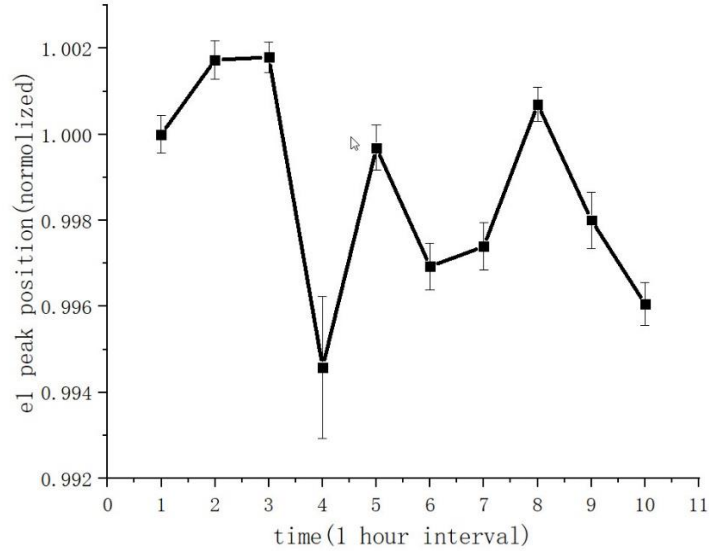

Figure 7. Variation of $e_1$ peak position with time.

## 4.2 Extraction of the intrinsic energy resolution

The simulated coincidence measurement of the energy deposition spectrum of the $^{207}$Bi source includes gamma rays and electrons in the liquid scintillator detector with energy selection in the coincidence detector, as shown in Figure 8a. By fitting the 976 keV energy peak, the contribution of the radioactive source was obtained as $\delta_{Edep}(\sigma/E) = 0.10\% \pm 0.00\%$.

Gaussian broadening was performed for the simulated energy spectrum and then aligned it with the experimental energy spectrum by the minimum Chi-square optimization method. The relationship between Chi-square and Gaussian broadening parameter (sigma) is shown in Figure 8b, when sigma is around 1.0, Chi-square reaches the minimum value. The comparison between the optimized simulated broadened energy spectrum and the experimental energy spectrum is shown in Figure 8c, the gray line is the experimental energy spectrum, the blue line is the simulated total energy spectrum, the green line is the $e_1$ energy spectrum, the light blue line is the $e_2$ energy spectrum, the black line is the gamma energy spectrum, and the red dashed line is the Gaussian fit of the peak position of $e_1$. The energy resolution of $e_1$ is 2.96% $\pm$ 0.01%, a bit more accurate than the double Gaussian fit. The peak position is shifted to the left by three thousandths (2.8 keV), mainly because the simulated peak of $e_1$ is not actually a strictly symmetric structure, but an anti-Landau asymmetric structure, which is due to the loss of electron energy in the air and scattering on the radiation source substrate.

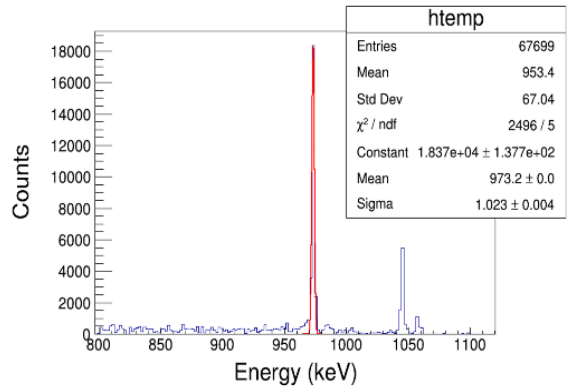

(a)

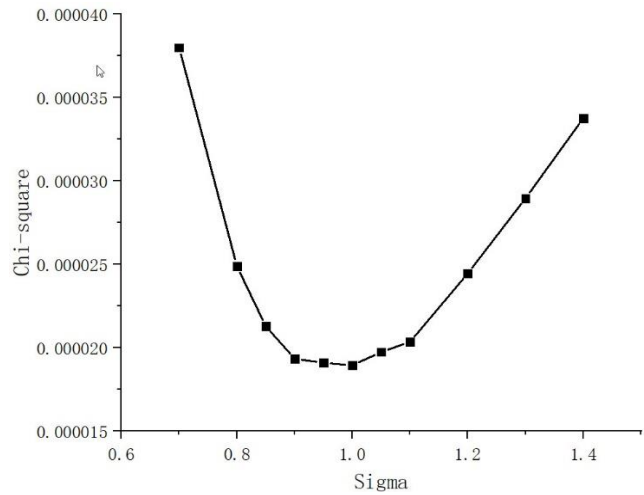

(b)

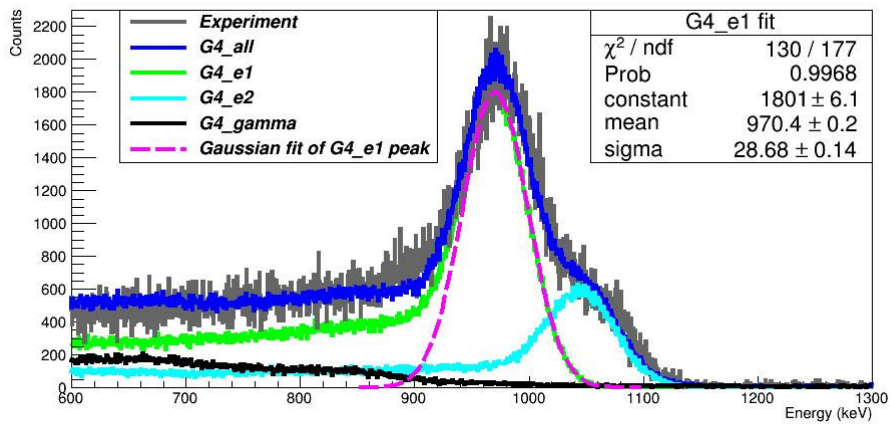

(c)

Figure 8. (a) Energy spectrum of the simulated radioactive source in the liquid scintillator. (b) Chi-square optimization for simulation energy spectrum Gaussian broadening. (c) Comparison of the optimized simulated broadened spectrum (blue) with the experimental spectrum (gray), and extracted $e_1$ (green), $e_2$ (light blue), gamma (black) spectrum and fit of $e_1$ peak (red dashed).

The measurement results for the horizontal uniformity of different light-emitting positions are shown in Figure 9a. Each point in the figure was obtained by fitting the peak position at 976 keV. The light output was reduced by approximately 3% near the edge of the container. Calculating the contribution of uniformity needs to consider the weight of the location of each measurement point. Therefore, the detector was divided into four concentric ring areas along the diameter direction, with respective diameters of 10, 20, 30, and 35 mm. The area of the region was used as the weight of the data points located in it. The contribution of horizontal uniformity after considering the weights was 0.89% ± 0.11%.

The variation of photon collection efficiency at different positions along the radial direction for five depth of 1 mm, 2 mm, 3 mm, 4 mm and 5 mm was simulated, as shown in Figure 9b. On the central axis of the container, at a depth of 1 mm, the photon collection efficiency is 88.2%, the fluctuation is 0.37% and the change with depth is less than 0.15% within 5 mm. The photon collection efficiency decreases from the center to the edge, and the reduction value at the edge is less than 5% compared with the center position, which is consistent with the horizontal experimental results. The average photon collection efficiency at each depth is shown in Figure 9c, with a variation of less than 0.5%. Simplified considering the uniform distribution of deposition energy at different depths, the contribution of the vertical uniformity of photon collection is 0.17% ± 0.01%.

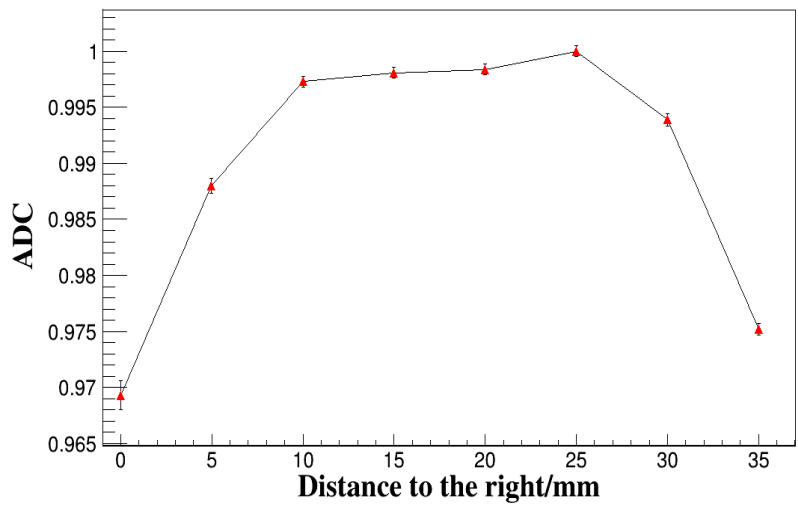

(a)

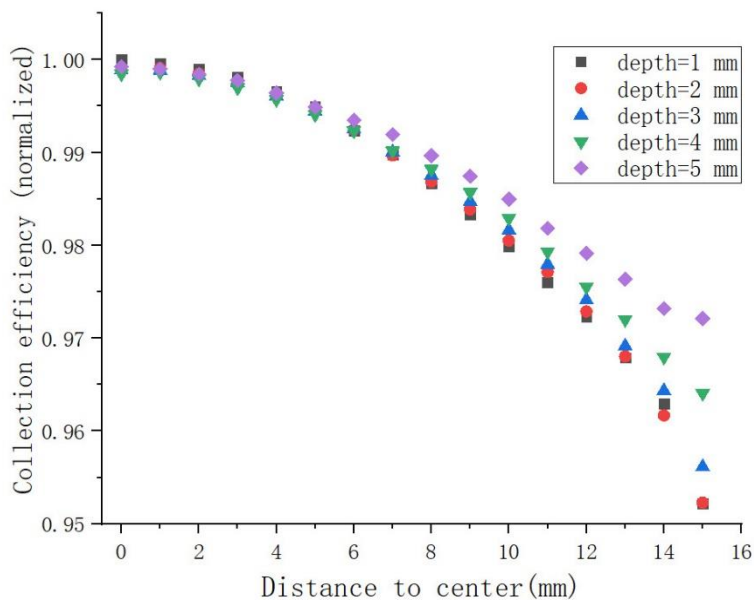

(b)

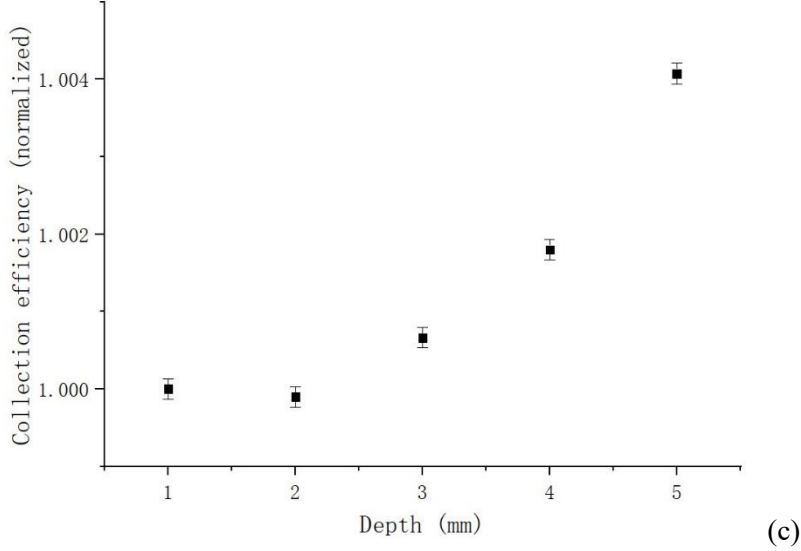

(c)

Figure 9. (a) The horizontal uniformity measurement of the liquid scintillator detector. (b) Photon collection efficiency at different depths as a function of distance to center. (c) Average photon collection efficiency at different depths.

The contribution of the test system was limited by the energy resolution measurement of the LaBr$_3$ crystal. Figure 10 shows the energy spectrum of the LaBr$_3$ crystal excited by $^{137}Cs$ radiation source. The energy resolution of 662 keV gamma ray is $(\sigma/E)_{LaBr3}$ =1.46%± 0.00%, and the corresponding photoelectron number was $N_{photoelectron}$= 11263 ± 0.0. Thus, by using these values, we calculated that the upper limit of the contribution of the measurement system $\delta_{system} \leq \sqrt{(\sigma/E)_{LaBr3}^2 - (\delta_{st_{LaBr3}})^2}$ = 1.19% ± 0.00%.

The resolution of single photoelectron ($(\sigma/E)_{sp.e.}$) was about 20% under different high voltages. The contribution of the fluctuation of the single photoelectron to the energy resolution is $\delta_{sp.e.} = (\sigma/E)_{sp.e.}/\sqrt{N_{photoelectrons}}$, according to the measured number of photoelectrons, liquid scintillator was 3266 and LaBr3 was 11263, its contribution to the energy resolution was 0.35% and 0.19%, respectively. After considering the contribution of single photoelectron, the upper limit of the system $\delta_{system}$ was revised to 1.17% ± 0.00%. Table 3 shows the results for the contribution of the energy resolution. Finally, the intrinsic energy resolution of the liquid scintillator is $\delta_{int}$ = 1.83% ± 0.06%.

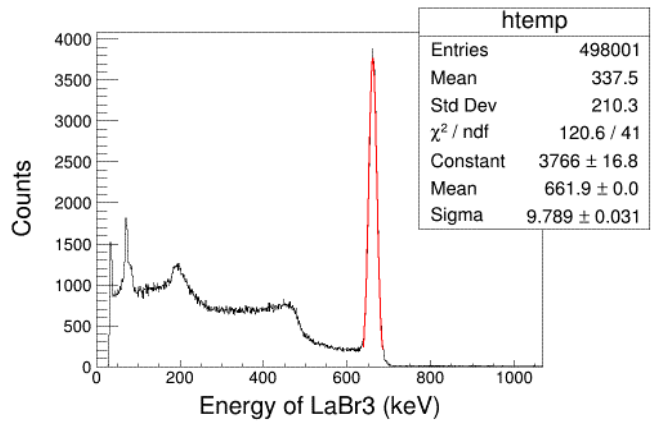

Figure 10. Energy spectrum of LaBr$_3$ crystal excited by $^{137}$Cs source.

Table 3. Experimental results of the contribution of energy resolution.

| Parameter | Formula | Fitted interval (keV) | Result |
|---|---|---|---|
| σ/E (a) | σ/E | 940-1110 | 2.96% ± 0.01% |
| $\delta_{st}$ (b) | $\sqrt{1/N_{photoelectrons}}$ | / | 1.75% ± 0.00% |
| $\delta_{system}$ (c) | $\sqrt{(\delta(\sigma/E)_{LaBr3})^2 - (\delta_{st_{LaBr3}})^2}$ | 640-685 | 1.17% ± 0.00% |
| $\delta_{Edep}$ (d) | σ/E | 960-980 | 0.10% ± 0.00% |
| $\delta_{uniformity}$ (e) | σ/mean | / | 0.89% ± 0.11%, 0.17% ± 0.01% |
| $\delta_{sp.e.}$ (f) | $(\sigma/E)_{sp.e.}/\sqrt{N_{photoelectrons}}$ | / | 0.35% ± 0.00% |
| $\delta_{stability}$ (g) | σ/mean | / | 0.16% ± 0.03% |
| $\delta_{int}$ | $\sqrt{a^2 - b^2 - c^2 - d^2 - e^2 - f^2 - g^2}$ | / | 1.83% ± 0.06% |

## 4.3 Light yield nonlinearity of liquid scintillator

The two-dimensional scatter plot obtained from the wide-angle Compton scattering experiment is shown in Figure 11. The X-axis represents the energy measured by HPGe detector, and the y-axis represents the ADC output of the liquid scintillator detector. The oblique line in the figure represents 662 keV scattering correlation events. The energy of HPGe detector was intercepted on the X-axis at a width of approximately 10 keV, and the scattering correlation events on the Y-axis were projected to obtain the ADC output of the liquid scintillator detector.

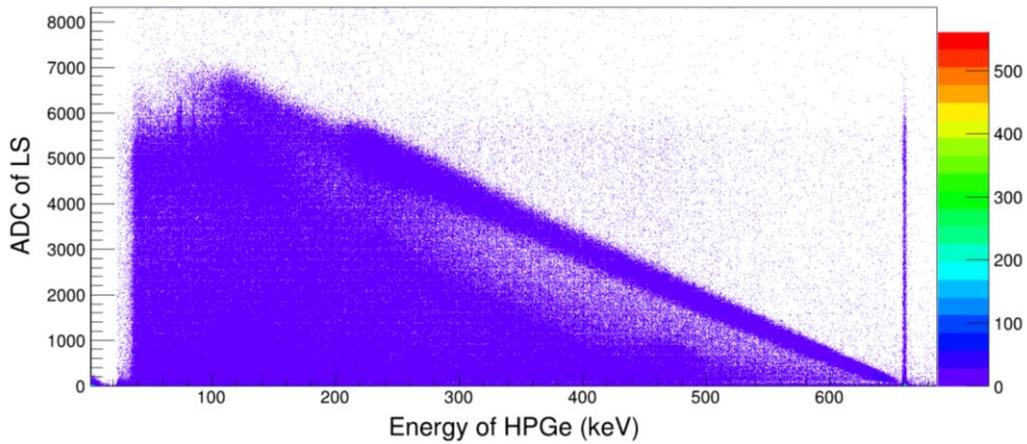

Figure 11. Two-dimensional plot of the energy spectrum of the HPGe detector and liquid scintillator detector.

The deposition energy ($E_e$) in the liquid scintillator corresponding to this ADC was 662 keV minus the energy measured by HPGe detector. Then, ADC/$E_e$ represents the light yield of the $E_e$ energy point. The curve normalized by the values of different energy points is the nonlinearity curve of the light yield, as shown in Figure 12. It is apparent that, the light yields above 350 keV tends to be stable, whereas that below 350 keV gradually decreases; the lower the energy, the faster the change and further decreases to 66% of 437 keV at 17 keV.

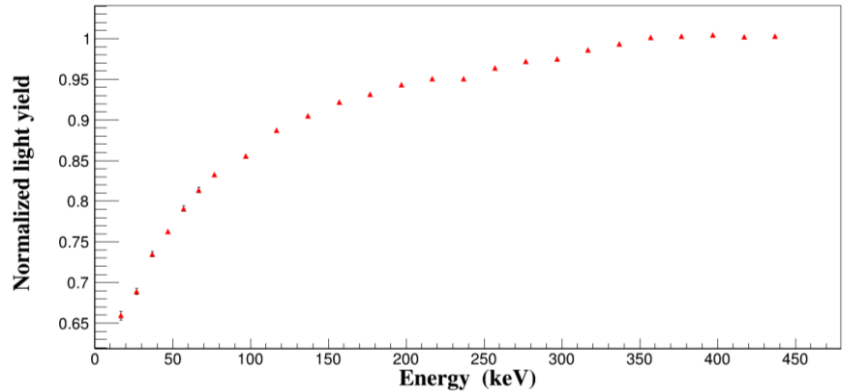

Figure 12. Light yield nonlinearity of liquid scintillator.

## 4.4 Energy spectrum reconstruction

The initial and final energies of each step were obtained using Geant4 to track the energy loss process of electrons in the liquid scintillator. The difference between the ADC corresponding to the energy at the beginning and the end of the step is the ADC output of this step. Adding the ADCs of all steps provides the total ADC output of one electron. Here the output value of ADC is the experimental value adopted at 17 - 437 keV, the fitted value adopted at 0 - 17 keV, and the linear assumption adopted at 437 - 1000 keV. The energy spectrum of the $^{207}$Bi source reconstructed using this method is shown in Figure 13. Fitting the energy spectrum by Gaussian can get the energy resolution as 0.4% ± 0.00%. This result demonstrates that the energy nonlinearity of the light yield does not contribute significantly to the intrinsic energy resolution.

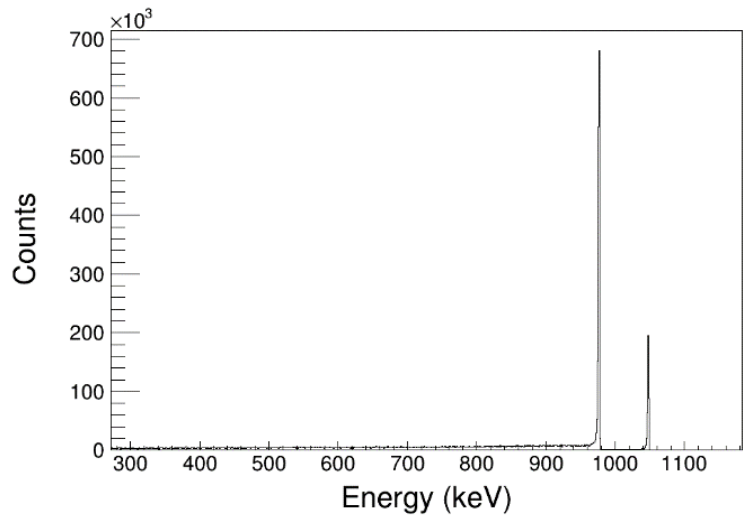

Figure 13. Reconstructed energy spectrum of $^{207}$Bi based on simulation and light yield nonlinearity of liquid scintillator.

## 5 Discussion and conclusion

A new method based on coincidence measurement and a $^{207}$Bi electron source for measuring the intrinsic energy resolution of 1 MeV electrons has been established in this study. Coincidence measurement can effectively suppress the influence of large-angle scattering events of gamma rays on single-energy electron energy spectrum measurements. By using a small liquid scintillator detector and a careful analysis of various factors contributing to the energy resolution, it was found that the intrinsic energy resolution of the liquid scintillator was 1.83% $\pm$ 0.06%. The subtraction of each resolution component from the total measured one using the square root was a conservative way. The correlation among these components was not studied yet.

The results of energy reconstruction based on simulation show that nonlinearity did not contribute significantly to the intrinsic resolution of electrons. Such an unexpected large intrinsic energy resolution may come from fluctuations in energy transfer processes. There may be large statistical fluctuations in the process from the energy loss of electrons to the emission of photons. The luminous efficiency of the liquid scintillator is relatively low, only approximately 3% of the deposited energy is converted to photons. This indicates that the energy transfer inside the liquid scintillator is relatively inefficient and may contribute to a relatively large statistical fluctuation, which requires further study.

# Acknowledgments

This research was supported by the National Natural Science Foundation of China (11775252). Thanks Liangjian Wen, Gaosong Li, Zeyuan Yu, Liang Zhan and Polao Lombardi for their detailed comments and suggestions, and special thanks Xiurong Li for her effective support.